\documentclass[useAMS,usenatbib]{mn2e}
\usepackage[final]{graphics}
\usepackage{amssymb}
\title[Discovery of H$\alpha$ satellite emission in a low state of the SW Sextantis star BB Doradus]{ 
Discovery of H$\alpha$ satellite emission in a low state of the SW Sextantis star BB Doradus
\thanks{Based on observations collected at the European Southern Observatory,
        Paranal (program 082.D-0154).}}
\author[L. Schmidtobreick, P. Rodriguez--Gil, et al.]
{L. Schmidtobreick$^{1}$\thanks{E-mail: lschmidt@eso.org},
 P. Rodr\'\i guez-Gil$^{2,3,4}$,
 K.\,S. Long$^{5}$,
 B.\,T. G\"ansicke$^{6}$,
\newauthor
 C. Tappert$^{7}$,
 M.~A.~P. Torres$^{8,9}$
\\
$^{1}$European Southern Observatory, Casilla 19001, Santiago 19, Chile\\
$^{2}$Instituto de Astrof\'\i sica de Canarias, V\'\i a L\'actea s/n, La Laguna,
 E-38205, Santa Cruz de Tenerife, Spain\\ 
$^{3}$Departamento de Astrof\'\i sica, Universidad de La Laguna, La Laguna, E-38203, Santa Cruz de Tenerife, Spain\\
$^{4}$Isaac Newton Group of Telescopes, Apartado de correos 321, E-38700 Santa Cruz de la Palma, Spain\\
$^{5}$Space Telescope Science Institute, 3700 San Martin Drive, Baltimore, MD 21218, USA\\
$^{6}$Department of Physics, University of Warwick, Coventry CV4 7AL, UK\\
$^{7}$Departamento de F\'\i sica y Astronom\'\i a, Universidad de Valpara\'\i so, Avda.\ Gran Breta\~na 1111, Valpara\'\i so, Chile\\
$^{8}$SRON, Netherlands Institute for Space Research, Sorbonnelaan 2, 3584 CA, Utrecht, The Netherlands\\
$^{9}$Harvard-Smithsonian Center for Astrophysics, 60 Garden Street, Cambridge, MA 02138, USA
}
 
\begin{document}

\date{xxxx}

\pagerange{\pageref{firstpage}--\pageref{lastpage}} \pubyear{2012}

\maketitle

\label{firstpage}

\begin{abstract}
BB\,Dor was observed during its low state state in 2009. Signatures of both 
binary components are revealed in the average optical spectrum;  
no signature of accretion is observed. Narrow emission lines
of H$\alpha$, He\,I and Na-D, as well as TiO absorption troughs
trace the motion of the irradiated
secondary star. We detect two additional components in the H$\alpha$
emission line that share many characteristics of similar "satellite" 
lines observed in the low state of magnetic cataclysmic variables of
AM\,Her type. It is the first time such emission components are detected
for an SW\,Sex star.
\end{abstract}

\begin{keywords}
Physical data and processes: accretion, accretion discs 
stars: activity -- fundamental parameters -- novae, cataclysmic variables  -- individual:BB Dor -- magnetic fields -- winds, outflows
\end{keywords}

\section{Introduction}
BB Doradus is a cataclysmic variable (CV) of the SW Sextantis type.
With typical periods between 3\,h and 4\,h these 
stars populate the upper edge of the period gap
and are believed to experience very high mass transfer rates. For a summary
on this subgroup, see the original paper by \cite{thorstensenetal91-1} and a more recent review by \cite{rodriguez-giletal07-1}.
Some of these stars show occasional low states when mass transfer is
reduced or even completely suppressed resulting in a brightness drop by 
$\sim 3-5$\,mag. The stars can stay at these low level for weeks or months
before rising again to their "normal" high state.
The advantage in observing the binary at low state lies in the reduced
mass transfer which thus results in a weak or absent accretion disc. While the
disc usually dominates the emission of a high mass transfer CV, in low state
one has a chance to observe signatures of the binary components: the
white dwarf primary and the secondary star. 
To our knowledge, the individual binary components could be observed for
only three systems in such a low state:
DW\,UMa \citep{araujo-betancoretal03-1,kniggeetal04-1}, TT\,Ari \citep{gaensickeetal99-1}, and MV\,Lyr \citep{2004ApJ...604..346H}.
In the context of an ongoing large program to study SW\,Sextantis stars 
(see e.g. \cite{2007MNRAS.374.1359R}), we monitor several of 
them photometrically to register when they enter one of these rare
low states and thus trigger detailed observations during this phase.
For more information on this project, see \citet{rodriguez-giletal11}.

\section{Observations and data reduction}
Time--resolved spectroscopic data were obtained 
on 2009 Jan 17 between 00:30 and 07:00\,UT when BB\,Dor was well in a
low state that lasted from May 2008 to April 2009 \citep{rodriguez-giletal12}. 
Target--of--Opportunity observations were triggered for the
FOcal Reducer and low dispersion Spectrograph (FORS2) \citep{appenzelleretal98-2}
on the Very Large Telescope (VLT) on Paranal, operated by the 
European Southern Observatory (ESO). We opted for GRIS\_1200R+93 and a 
$0.7''$ slit to concentrate on the H$\alpha$ emission line. Clear skies 
and a stable seeing around $0.8''$ allowed to observe
65 spectra with an individual exposure time of 300\,s. In total,
we covered 6.5\,h, about 1.5 orbital cycles of BB\,Dor.

The data were reduced using standard procedures in 
{\tt IRAF}\footnote{{\tt IRAF} is distributed by the National Optical Astronomy Observatories}. 
Further analysis was done using 
{\tt MOLLY}\footnote{Tom Marsh's package {\tt MOLLY} is available at 
http://deneb.astro.warwick.ac.uk/phsaap/software/}, 
{\tt MIDAS}\footnote{{\tt MIDAS} is distributed by the European Southern 
Observatory at http://www.eso.org/sci/ \mbox{software/esomidas/}}, 
and self-written C-routines.
The spectra were
optimally extracted following the method by \citet{horne86-1},
the final Full-Width-at-Half-Maximum (FWHM) resolution is 2.05\,\AA.
We used standard star observations 
to flux-calibrate the spectra. The night was classified as clear but not 
photometric so thin clouds might have been passing occasionally. Therefore, 
we do not claim the absolute flux values to be accurate by more than 0.5\,mag.
However, the
flux-calibration removes the detector specific spectral response and yields
relative flux values that can be compared within the average spectrum. 
\section{Results}
\subsection{The average spectrum}
\begin{figure}
\rotatebox{270}{\resizebox{!}{8.6cm}{\includegraphics{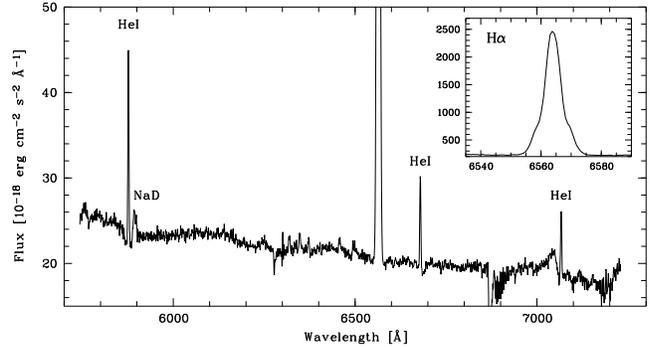}}}
 \caption{\label{average_vlt}
The average of our 65 spectra. The main emission lines are labelled. In the upper right corner a zoom on H$\alpha$ is given.}
\end{figure}

In Fig.\ \ref{average_vlt}, the average of the 65 spectra is plotted. It is
dominated by narrow emission lines of hydrogen and helium. 
The FWHM of H$\alpha$ is 6.9\,\AA, the He\,I lines have 3.4\,\AA\ on average.
\citet{rodriguez-giletal12} show that accretion events in BB\,Dor during 
this low state happen rarely. These epochs can be clearly distinguished 
from those without accretion as the spectra show broader
emission line profiles and a bluer continuum when accretion happens.
We compared the line profiles and continuum in the individual spectra 
with those in
\citet{rodriguez-giletal12} and find no indication of an accretion event in 
our data. Indeed,
the widths of the lines are narrow and very stable throughout the 
observations as 
expected for a non--accreting CV where the origin of such emission lies 
in the chromosphere or irradiated surface of the companion. 
This is also confirmed later in this paper
by our study of radial velocities.

Around the H$\alpha$ and He\,I emission lines one can
detect weak but broad absorption features. Their origin is most certainly the
white dwarf which has been accreting until recently and thus is expected
to have He present in its atmosphere. The presence of He\,I 
rather than He\,II
suggests a white dwarf temperature between 20000\,K and 40000\,K
which is rather low compared to other SW\,Sex type stars \citep[MV\,Lyr: 47000\,K, DW\,UMa: 50000\,K, TT\,Ari: 39000\,K,][]{Town+09}.
Still, it is 
in agreement with the value $T=37000$\,K found from Far UV spectroscopy of
BB Dor in high state \citep{godonetal08-1} and a value between 25000\,K and
30000\,K
found by 
\citet{rodriguez-giletal12} from a WD+M\,dwarf composite fit to a spectrum 
covering the whole visual spectral range of BB\,Dor in quiescence. 

We checked the absorption lines for the presence of 
Zeeman-splitting to get an estimation for a possible magnetic 
field on the white dwarf. No splitting was detected in any of these lines. 
From this we infer an upper limit of
the magnetic field to about 5\,MG.

The secondary star shows itself through
the presence of weak TiO absorption troughs.

\subsection{The H$\alpha$ emission distribution}
Radial velocities were measured by fitting Gaussians of 
300~km~s$^{-1}$
width to the core of the H$\alpha$ emission lines. Using the 
analysis--of--variance algorithm \citep{schwarzenberg-czerny89-1} as
implemented in MIDAS
we found the most likely orbital period $P=0.149(4)$\,days. This value 
is in agreement with \cite{rodriguez-giletal12} 
and places BB\,Dor right at the centre of the period distribution
of the SW\,Sex stars. By combining their multi-epoch data with our data
Rodr\'\i guez-Gil et al.\ derive a much more precise value 
for the orbital period of $P=0.154095(3)$\,days. We adopt this latter
value for all our further analysis, and also use their zero-phase
$T_0({\rm HJD})=2454833.7779\pm 0.0003$ defined as the blue-to-red crossing
of the H$\alpha$ velocities.

\begin{figure}
\parbox[b]{4.35cm}{
\rotatebox{-90}{\resizebox{!}{4.35cm}{\includegraphics{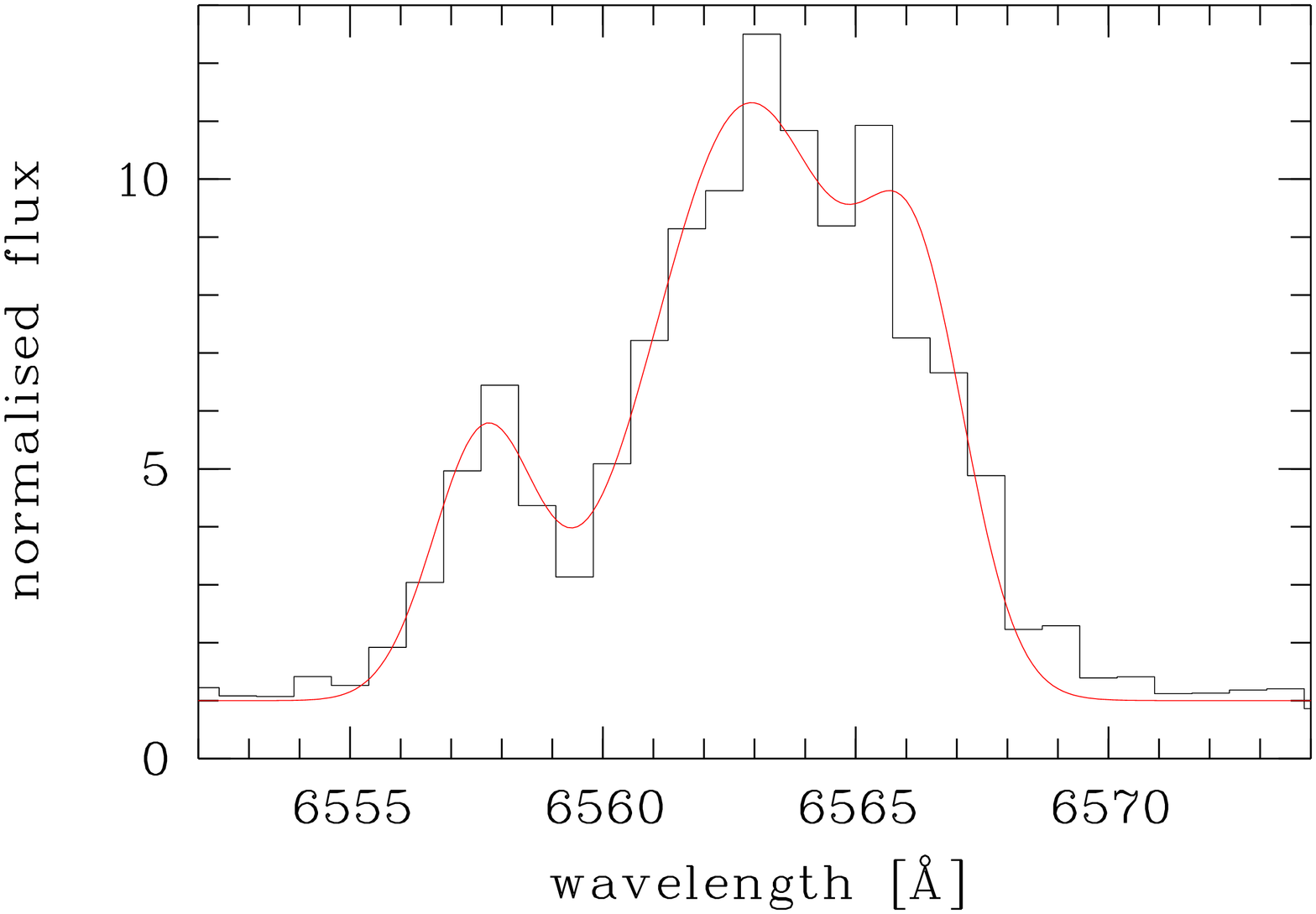}}}}
\hfill
\parbox[b]{3.9cm}{
\rotatebox{-90}{\resizebox{!}{3.85cm}{\includegraphics{ha_bbdor_vlt_trailed_core.ps}}}}
 \caption{\label{ha_satellites}
On the left, a zoom on the H$\alpha$ line of an example spectrum 
(orbital phase 0.93) is plotted with a 3--Gaussian fit for the 
central line and two satellite lines. 
The trailed spectra diagram on the right shows the orbital variation 
of the line profile.}
\end{figure}

Studying the H$\alpha$ emission line in more detail reveals a remarkable
structure:
Its line profile
consists of at least three components, a central emission line and two satellite
lines (see zoom in Fig.\ \ref{ha_satellites}). 
Their orbital variation is demonstrated 
in the plot on the right: the trailed spectra diagram of H$\alpha$.
The two satellite lines 
can be clearly distinguished. They have the same radial velocity amplitude of 340\,km\,s$^{-1}$
which is larger than that of the central line. They are
symmetrically offset in phase by $\pm 0.15(2)$. 

\begin{figure*}
a)\resizebox{!}{3.7cm}{\includegraphics{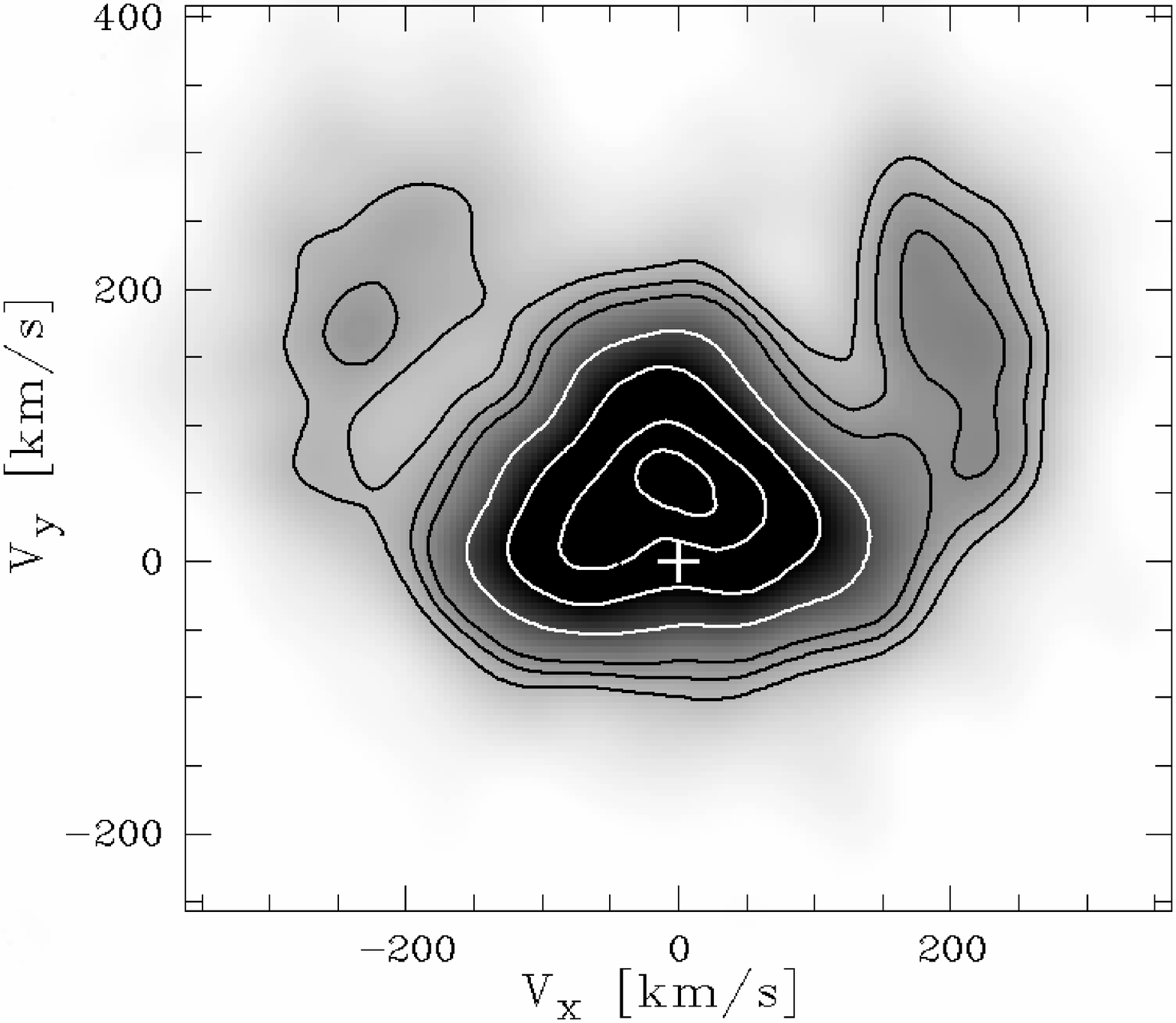}}
\hfill
b)\resizebox{!}{3.7cm}{\includegraphics{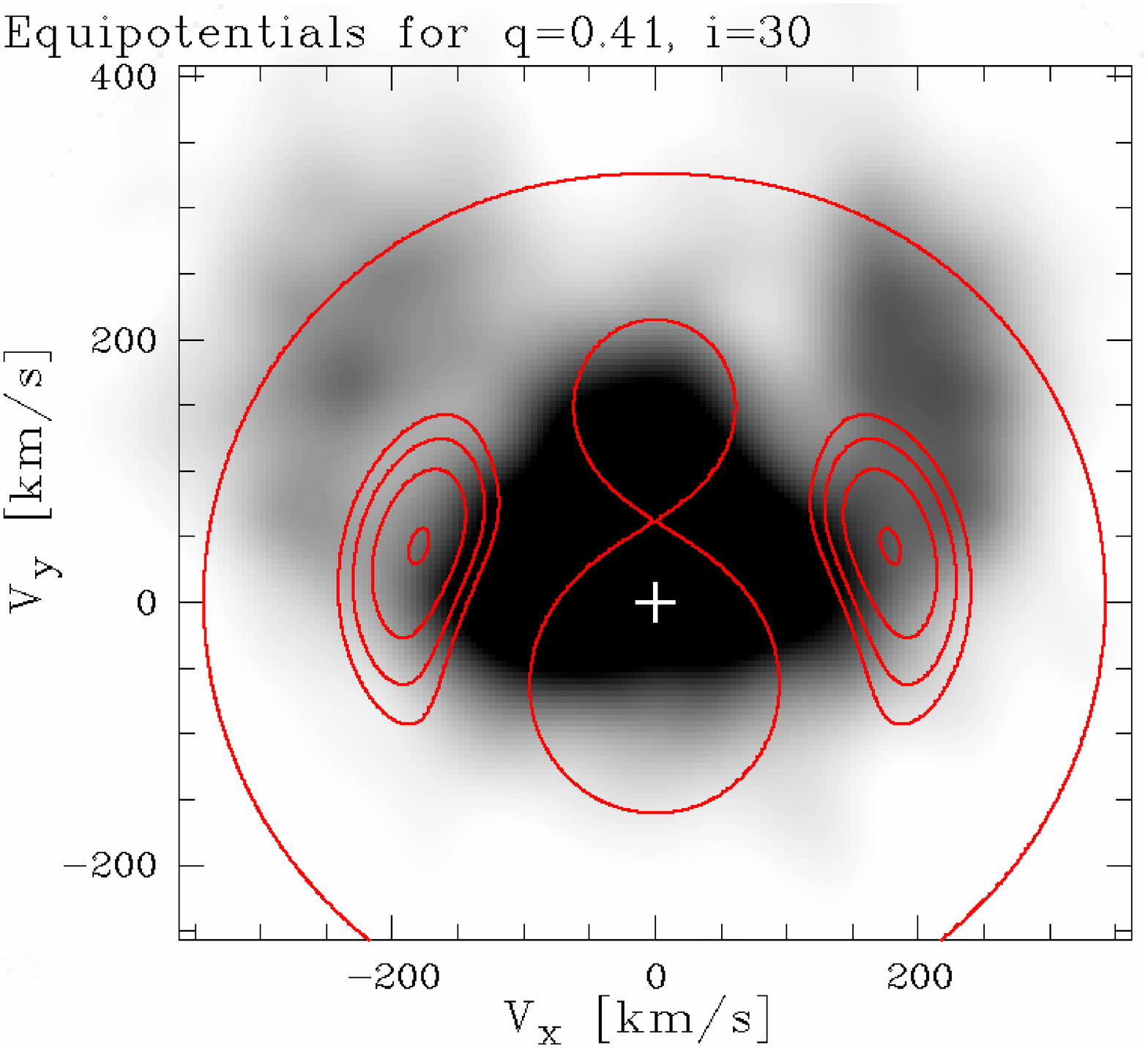}}
\hfill
c)\resizebox{!}{3.7cm}{\includegraphics{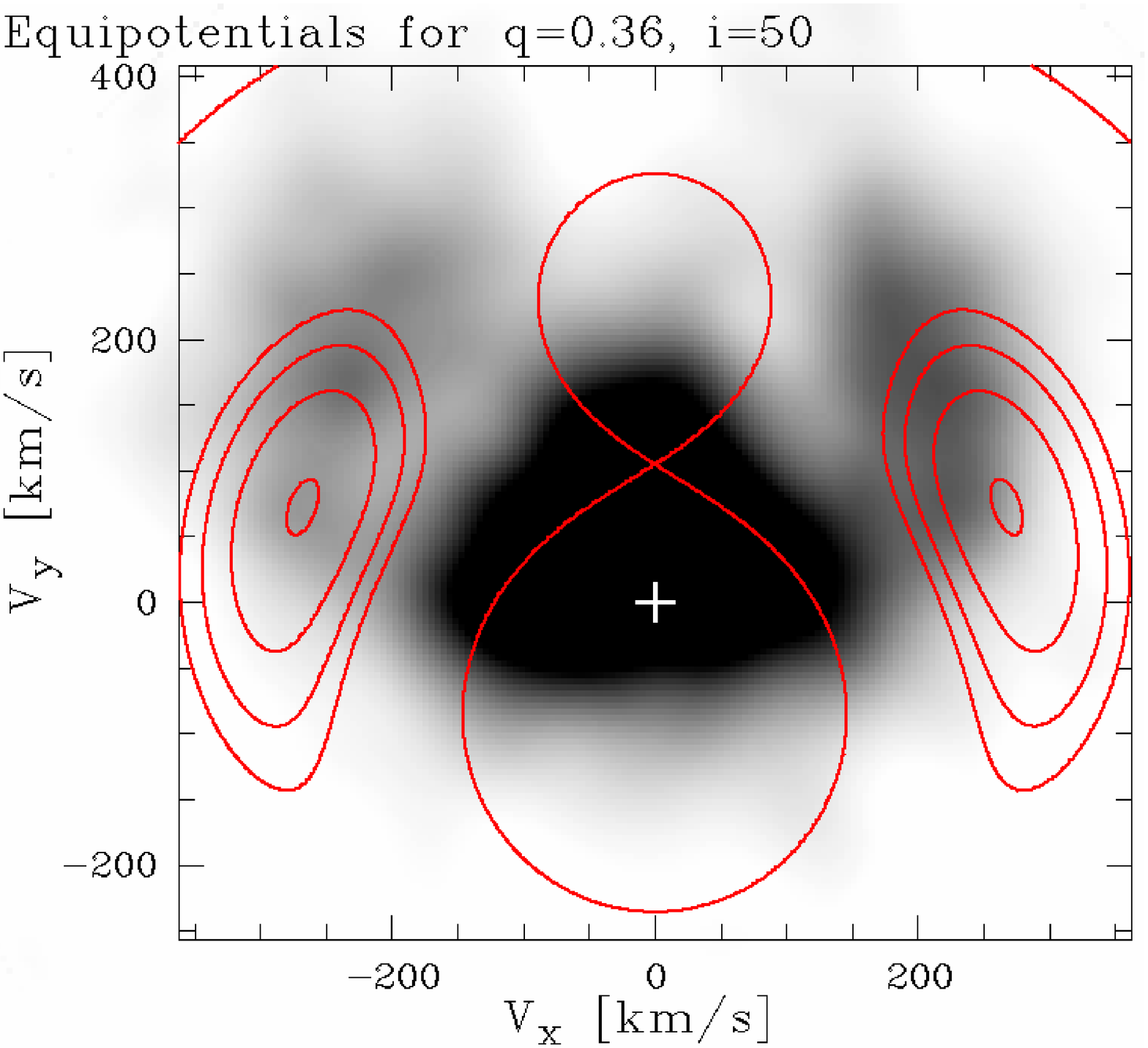}}
\hfill
d)\resizebox{!}{3.7cm}{\includegraphics{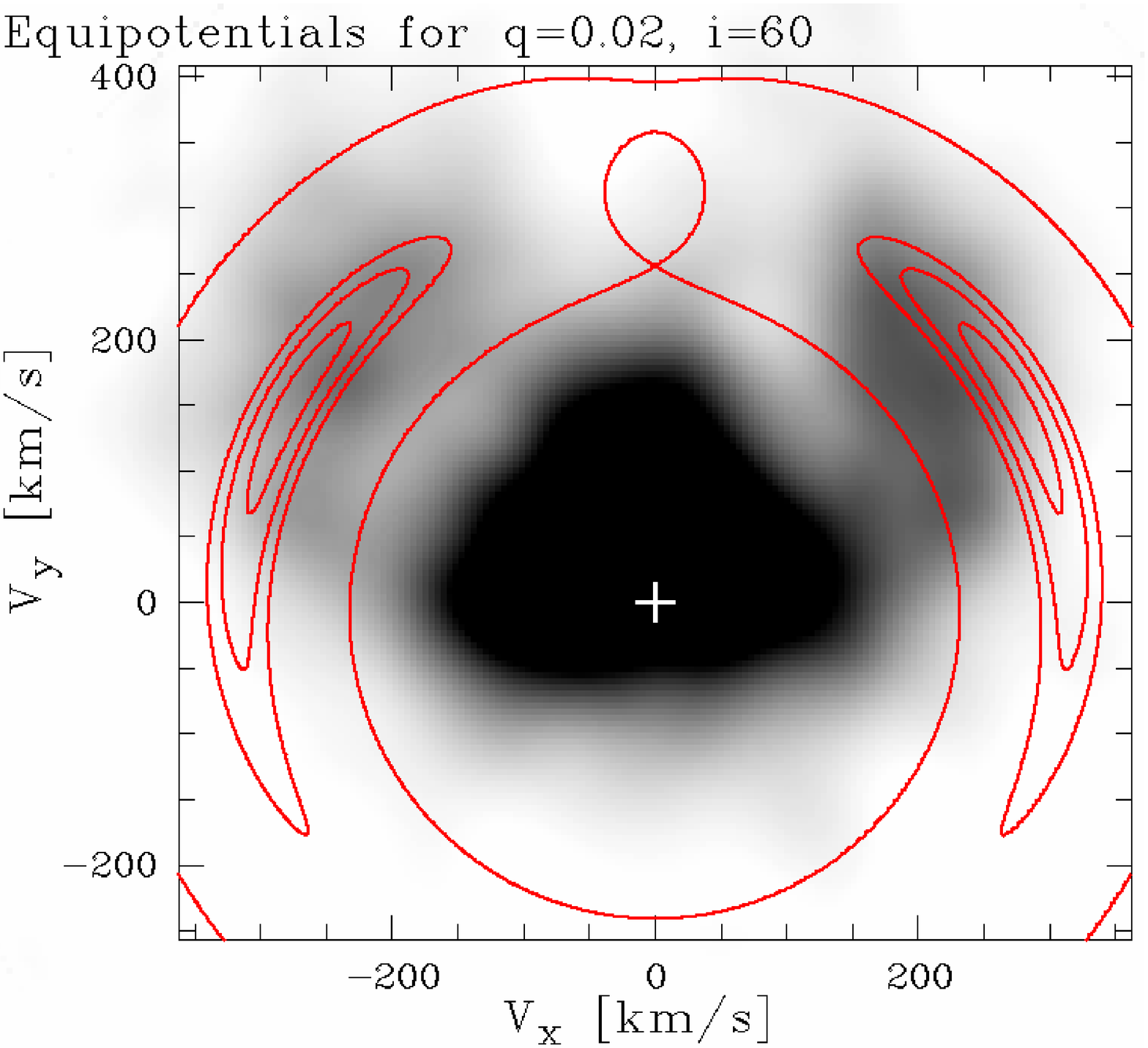}}
\caption{\label{dop_ha}
Doppler map of the H$\alpha$ emission distribution in BB\,Dor.
On the left side, the contours of the emission distribution are overplotted.
The three images to the 
right have the Roche equipotentials overplotted for different
mass ratios and inclinations, see text for details. The centre of mass is given as $+$.
No match using reasonable parameters can be found between the
symmetrical, extended structure and the outer Lagrangian points}
\end{figure*}

These are the first ever observations of such satellite lines 
in an SW\,Sex star. However, such satellite lines 
were detected before in the quiescent states of polars, i.e.
strongly magnetic CVs \citep{kafkaetal05-1, kafkaetal06-1, 2007AJ....133.1645K, kafkaetal08-1, 2010ApJ...721.1714K, 2008A&A...490..279M}. 
First hints for 
such an additional emission are also present in earlier low resolution data
of AM\,Her \citep{lathametal81-1}. We also find indications for
a similar feature
in the low-state spectra of the non-magnetic novalike TT\,Ari that were,
however, interpreted differently at the time \citep{shafteretal85-1}:
In their Figure~10, a small line component is visibly moving around the
central H$\alpha$ line. The radial velocity of this component is given in 
their Figure~13.  While it is fitted there with a single sinusoid, the
data could in retrospect also be interpreted as two crossing sinusoids 
as we observe them in BB\,Dor. In the detached binary V471\,Tau,
\citet{youngetal91-1} observe such additional line components and
interpret them as evidence for extended clouds or a ring of plasma orbiting
the K-dwarf secondary. All these observations indicate that the here
presented feature of satellite lines might be common to a number of CVs and 
related objects and is not necessarily limited to magnetic ones.

Comparing our data to the well-observed AM\,Her stars, the appearance of 
these satellite lines is strikingly similar.
\citet{kafkaetal08-1} discuss
partial loops of streaming gas co-rotating with the donor star or collimated
jet-like outflows as possible origin for these satellite lines. However, the
interpretation remains vague due to the uncertainty whether the lines follow
crossing or parallel sinusoids.
In the case of AM\,Her, the lines were not sufficiently 
resolved for this distinction,
while the other data are of even poorer quality or 
do not cover a complete orbit. Only for BL\,Hyi, \citet{2010ApJ...721.1714K}
conclude that crossing satellite lines are most likely to explain their data.

For BB\,Dor, one can distinguish the three components over the 
full orbit (compare Fig.\ \ref{ha_satellites}) and thus answer this question:
We clearly see that a crossing of the two lines
occurs at phases 0.25 (red side) and 0.75 (blue side). We also point out that
the phase offsets of the satellite lines are extremely symmetrical with 
respect to the phase of superior conjunction of the 
secondary. And more, if we assume that the
satellite lines in the AM\,Her stars are also described by crossing lines, the
phase offsets there show the same symmetry. Having 
perfect symmetry in all cases so far, seems to indicate rather a rule than a 
coincidence. Therefore, assuming that the satellite lines observed in BB\,Dor
and in the polars
are stable and not just a transient feature, any explanation for the 
presence of these lines 
also has to explain their symmetry. 
An origin in random loops or prominences on the companion would not yield such an explanation
and is thus refuted. 
It is rather more likely that
the material in which the satellite lines are emitted is confined in
two locations which are favoured by the force field of the system.
Assuming that only the gravitational force is acting in the binary,
this would e.g. support the idea that the material is trapped in the L4 and L5
Lagrangian points 
which was already discussed by \cite{kafkaetal08-1} for 
AM\,Her. They argued against this idea because the L4 and L5 points only 
yield a stable equilibrium for mass ratios of $q \lesssim 0.04$ which 
are physically implausible for AM\,Her.
However, one can deem similar points of equilibrium possible
if the Roche geometry is modified by additional forces, e.g. 
the magnetic field of any of the two stars. Note that due to the general 
nature of the magnetic field, these equilibrium points are most likely not
situated in the orbital plane of the system even though they would be bound to 
the orbital motion. 

As a first test on the origin of the satellite lines, we 
computed a Doppler--tomogram of the H$\alpha$ emission in BB\,Dor
using the code by \citet{spru98} with a {\sc MIDAS} interface \citep{tappertetal03-1}.
The resulting map is plotted in Fig.\,\ref{dop_ha}\,a), the 
satellite lines are clearly visible as two arcs reaching out
from the main triangular-shaped emission region close
to the centre of mass towards positive $v_y$. In b)-d), we 
overplotted
the Roche-geometry for various inclinations and mass ratios. 
Since the inclination of BB\,Dor is unknown but considered
on the lower side \citep{2001MNRAS.325...89C} we varied it arbitrary between 30$^\circ$
and 60$^\circ$.
We used the sole dependence of a Roche-lobe-filling secondary's density 
with the orbital period and the mass/radius relation for low mass red dwarfs
$R=M^{0.0867}$ \citep{hellier_book} to derive $M_2 = 0.33 M_\odot$. With
an average white dwarf mass of $M_1 = 0.8\,M_\odot$ \citep{2011A&A...536A..42Z} we calculate the mass ratio
$q = 0.41$ (Fig.\,\ref{dop_ha}\,b).
The plot in Fig.\,\ref{dop_ha}\,c) shows the results for the
mass ratio $q = 0.36$ based on the revised 
evolutionary track by \citet{2011ApJS..194...28K}, while
Fig.\,\ref{dop_ha}\,d) was calculated to force a match of the two
arcs with the Lagrangian points 
L4 and L5. 
For low inclinations ($30^\circ \le i \le 40^\circ$) and reasonable mass ratios, 
the L4 and L5 points fall together with
the corners of the inner triangular structure. This might hint at an outflow
of material through the Lagrangian points L4 and L5. Since the 
agreement depends strongly on the selection of the mass ratio and 
the inclination, these values need to be better constrained to support
such a claim.

For none of any 
-- even generously interpreted-- possible 
values for the mass ratio of a system just
above the period gap ($0.2< M_2/M_1 <0.6$), a kinematic match is found 
between the elongated structure and the L4 and L5 points.
In fact, to force such a match,
a mass ratio of about 0.02 is needed which is far 
away from any physically reasonable value for a system with an orbital 
period of about 3.7\,h. An explanation through pure Roche potential is
thus not possible. This is in agreement with \citet{kafkaetal08-1} 
who ruled out an explanation through pure Roche potential
for stability reasons. 
Just for completeness, we note that
the necessary mass ratio of $q = 0.02$ does actually yield a stable equilibrium 
at the L4 and L5 points as it is below the upper limit of 0.04. 
Again, since this critical value for the mass ratio of 0.04
is highly exceeded by systems above the period gap, they can not
have a pure Roche potential with the L4 and L5 points being stable
equilibrium points.
One may speculate, however,
whether a Roche 
geometry where the force field is modified by including a suitable 
magnetic field
could actually result in the same stable equilibrium points 
around positions L4$_{\rm mod}$ and L5$_{\rm mod}$ as if the 
force field was purely gravitational but resulting from a
low mass ratio. For binaries, such modifications of the Roche potential were
performed by \citet{2009AA...507..891D} to account
for additional forces due to radiation pressure and pulsation.
For single massive stars with large dipole magnetic fields, points of 
equilibrium between the gravitational, centrifugal and magnetic
forces were modelled by \citet{2005MNRAS.357..251T} and observed by 
\citet{2011MNRAS.tmp.1620O}. It thus seems natural that
similar points can also exist in close binary systems.
\begin{figure}
\resizebox{5.4cm}{!}{\includegraphics{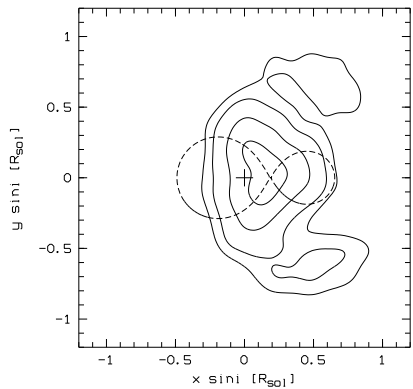}}
 \caption{\label{plot_xy}The H$\alpha$ emission distribution is plotted
as solid contours in
Cartesian coordinates that were calculated with the assumption
that all emission sources are bound to the orbit and have no individual
velocities. Overplotted in dashed lines is the Roche-geometry for
$q = 0.41$ and $i=30^\circ$.}
\end{figure}

This could then naturally explain the distribution of the material observed here.
In this picture, the presence of similar satellite lines in other close 
binary systems would require the right combination of gravitational and 
magnetic forces to yield stable equilibrium points around which the material
could be confined. This might explain 
why these satellite lines are not observed in all systems. 
Since in this scenario, the emission sources are co-rotating with the secondary
their radial velocities can easily be converted into space coordinates 
using the orbital period.
In Fig.\ \ref{plot_xy} we have done this and for comparison overplotted
the Roche lobe for $q=0.41$ and $i=30^\circ$. 

On the other hand, the comparison with the Roche geometry only makes sense 
for those emission 
regions that are moving along with the orbit of the binary. Any 
velocity component vertical to the orbital plane will add to 
the radial velocity component within the orbital plane the  
radial component of this vertical velocity, which is constant over
the orbital period as long as the orbit itself does not vary in space. 
The final result of such a vertical motion is thus a shift to the blue or
red of the observed line and the effect on 
the Doppler map is similar to the one obtained if the system velocity is 
not correctly applied. However, this is not what we observe for the H$\alpha$ 
satellite lines whose zero velocity agrees with the overall system velocity
as is seen in the trailed
spectra diagram (Fig.\ \ref{ha_satellites}). We can thus conclude that 
the satellite lines do not have a mayor vertical velocity component but
their velocity vectors are confined in the orbital plane. Note that
this does not imply that the material itself is actually situated in 
the orbital plane. It can be at any distance as long as it is moving 
parallel to the plane.

\begin{figure}
{\resizebox{8.4cm}{!}{\includegraphics{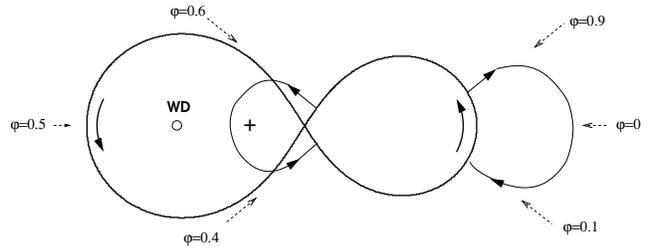}}}
 \caption{\label{lobes}Sketch to illustrate the two possible orientations
of prominences that could explain the observed satellite lines. Note that
the scaling and in particular the size of the lobes are arbitrary. }
\end{figure}

Since we observe the crossing of the two satellite lines, 
we can specifically exclude any
prominences or outflows (as discussed by \citet{kafkaetal08-1} for AM Her)
that move out of the orbital plane as their
origin. 
Any such prominence with a considerable vertical velocity 
component would result in shifted emission lines that move parallel 
to each other: their radial velocity amplitude thus belongs to the component
within the orbital plane while the shift is given by the radial velocity of the
vertical component. 
If the observed satellite lines are therefore due to prominences, these 
prominences have to be orientated in the following way: (1) they have to 
run within the orbital plane without a major vertical component and
(2) they have to be symmetrical with respect to the axis through the white 
dwarf and the secondary star. To account for the maximum radial velocity 
being observed at phases 0.1 and 0.4, the minimum values at 0.6 and 0.9,
the angle under which the material leaves the secondary must be 
$36 \pm 7^{\circ}$ on the preceding side, 
the angle at which it enters back must be $36 \pm 7^{\circ}$ on 
the recessing side.
This leaves only two possible orientations for the prominences that are 
illustrated in Fig.\ \ref{lobes}. We calculated the additional velocities
of the prominences that would be needed to yield the observed radial 
velocity variation. Depending 
on the size of the secondary and the orientation of the prominence, we obtain
between 200 and 350\,km\,s$^{-1}$. These values are in the range of 
the velocities observed for coronal mass ejections from the Sun
whose velocity distribution shows a strong peak
around 350 \,km\,s$^{-1}$ \citep{2005ApJ...619..599Y}.

Large prominences of similar velocities are observed in fast rotating
single stars like BO\,Mic and AB\,Dor. Except for the symmetry which is 
not present in single stars the spectral appearance is similar to
what we observe in BB\,Dor:
narrow emission lines move around the stellar disc and have radial 
velocity amplitudes up to $\sim$550\,km\,s$^{-1}$
\citep[see e.g.][]{2006MNRAS.373.1308D}.
For binaries, evidence for prominences comes from the X-rays.
\citet{jeffries96-1} model
X-ray spectra of XY\,UMa and find -- also in
comparison with other binaries \citep{hall+ramcey92-1} -- 
that large prominences
situated in the orbital plane are an apparent 
configuration in interacting binaries.
In fact, \cite{1996MNRAS.281..626S}
argue that stable magnetic loops would exist in the region between L1 and
the white dwarf making the presence of large slingshot prominences in this
part likely.

\subsection{Study of radial velocities}
\begin{figure}
\rotatebox{-90}{\resizebox{!}{8.4cm}{\includegraphics{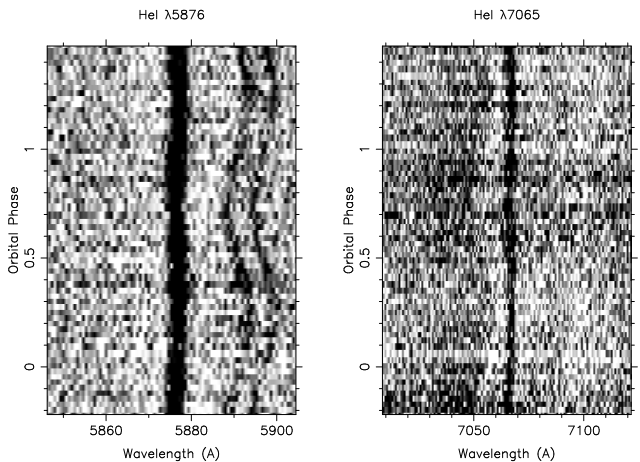}}}
 \caption{\label{hei5876_7065}
Left side: trailed spectra of He\,I\,$\lambda 5876$ and Na\,D. 
The difference in the amplitude of the radial velocities is clearly visible.
Right side: trailed spectra of He\,I\,$\lambda7065$. The emission line 
is seated inside a broad TiO 
absorption band that similar to Na\,I clearly moves with a higher
velocity than the He\,I emission line.}
\end{figure}

A detailed examination reveals
that all emission lines move in phase with the secondary.
In Fig.\ \ref{hei5876_7065}, the trailed spectra 
diagrams are plotted for the regions around He\,I\,$\lambda5876$ and He\,I\,$\lambda 7065$. It shows
the low velocity variation of the He\,I emission and in the left plot also 
the variation of the Na\,D doublet emission which is in phase 
with He and H$\alpha$ but displays a larger 
amplitude of the radial velocities. 
In the trailed spectra around He\,I\,$\lambda 7065$ on the right side,
a broad TiO absorption trough mimics the 
motion of the Na\,D lines but has an even larger velocity amplitude. 

\begin{figure}
\rotatebox{-90}{\resizebox{!}{8.4cm}{\includegraphics{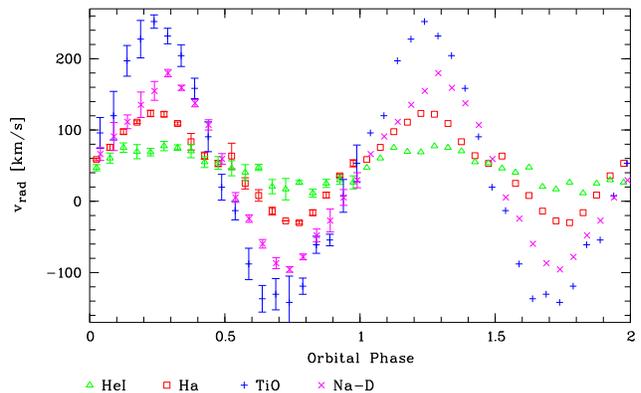}}}
 \caption{\label{plot_rv} Phase--binned radial velocities are plotted for the
H$\alpha$, He\,I, and Na\,D emission lines and for a TiO absorption trough.
}
\end{figure}

\begin{table}
\centering
\caption{\label{fit_par}The parameters of the sine fit 
$\gamma + k \cdot sin{2\pi (\phi -\phi_0)}$ to the radial velocities}
 \begin{tabular}{@{}cccc@{}}
 \hline
          & $\phi_0$ & $\gamma ~\rm [km\,s^{-1}$] & $k~\rm [km\,s^{-1}$]\\
 \hline
He\,I     & 0.023(07) & 46.8(9) & 28.5(12) \\
H$\alpha$ & -0.010(03) & 48.8(8) & 67.8(11) \\
Na\,D     & -0.013(03) & 41(2) & 124(2) \\
TiO       & -0.040(03) & na   & 186(3) \\
 \hline
\end{tabular}
\end{table}

Quantitatively, these results are supported by the phase folded radial velocity curves
plotted in Fig.\ \ref{plot_rv}. For the He\,I\,$\lambda$5878, He\,I\,$\lambda$6678, and 
He\,I\,$\lambda$7065 emission lines, the radial velocities were measured by 
fitting single Gaussians of 300\,km\,s$^{-1}$ width to the lines. All three lines
follow the same variation. However, He\,I\,$\lambda$7065 is very noisy, the line 
is not always detected in the single spectra so we decided to only use
He\,I\,$\lambda$5878 and He\,I\,$\lambda$6678 and averaged them within 
each phase bin. 
The Na\,D emission lines and the TiO absorption bands
were generally too weak to be measured in the individual spectra. 
Instead, we averaged 
data from the same orbital phase without changing the time--resolution and 
plotted them as trailed spectra. For TiO, we manually followed the position
of the edge of the absorption trough at 7052\,\AA\ on these trailed spectra.
We independently determined its radial velocities three times.
For Na\,D, we manually measured the radial velocities of each line twice. This
gave four independent measurements which were then phase binned. 
For all methods,
the error is determined as the sigma of the distribution of the radial
velocities within the bin size of 0.05 phases.
To confirm the manual measurement of the radial velocities by a more 
robust technique, we 
subtracted velocity sine waves with amplitudes ranging from
115 to 125 $\rm km\,s^{-1}$ to the 65 spectra to eliminate the motion of
the Na\,D lines. Radial velocity curves of the almost-straight Na\,I
lines were constructed and their deviation from the mean
measured. We then fitted a parabola to the velocity amplitude--sigma
data and measured its minimum as the correct velocity amplitude
$K_2 = 118 \pm 4 \,\rm km\,s^{-1}$.This value is in agreement 
with the one found by the manual method described above 
(see Table \ref{fit_par}).

Monte Carlo sine fits to the data (for the
resulting parameters, see Table \ref{fit_par}) 
show that the phase shifts of all 
radial velocity curves are in agreement with the average value  
indicating that all analysed
lines have their origin on the side of the secondary star. 
The system velocity $\gamma$
is determined as $48\pm 2$\,km\,s$^{-1}$.
As easily seen in 
Fig.\ \ref{plot_rv}, the amplitude of the radial velocity varies, 
however, significantly for different lines; 
the exact values are given in Table \ref{fit_par}. It is to note that for 
H$\alpha$ our value
$k = 67.8 \pm 1.1 \rm\,km\,s^{-1}$ differs 
significantly from the one
derived by \citet{rodriguez-giletal12} $k = 99 \pm 2 \rm\,km\,s^{-1}$.
This is probably due to the lower spectral resolution of 
\citet{rodriguez-giletal12} such that they do not resolve the satellite 
lines which will then contribute to the overall radial velocity. As a test,
we smeared our spectra with a Gaussian of 5.4\,\AA\ and
measured the radial velocities of H$\alpha$ with a Gaussian of 400km\,s$^{-1}$
thus including the satellite lines and simulating the data and measurements of \citet{rodriguez-giletal12}.
From these we obtain a value $k = 79 \pm 1 \rm\,km\,s^{-1}$ which is 
in slightly better agreement with \citet{rodriguez-giletal12} and 
thus confirms 
that their radial velocity amplitude is at least partly increased due to 
the unresolved component of the satellite lines. 
\begin{figure}
\resizebox{4.1cm}{!}{\includegraphics{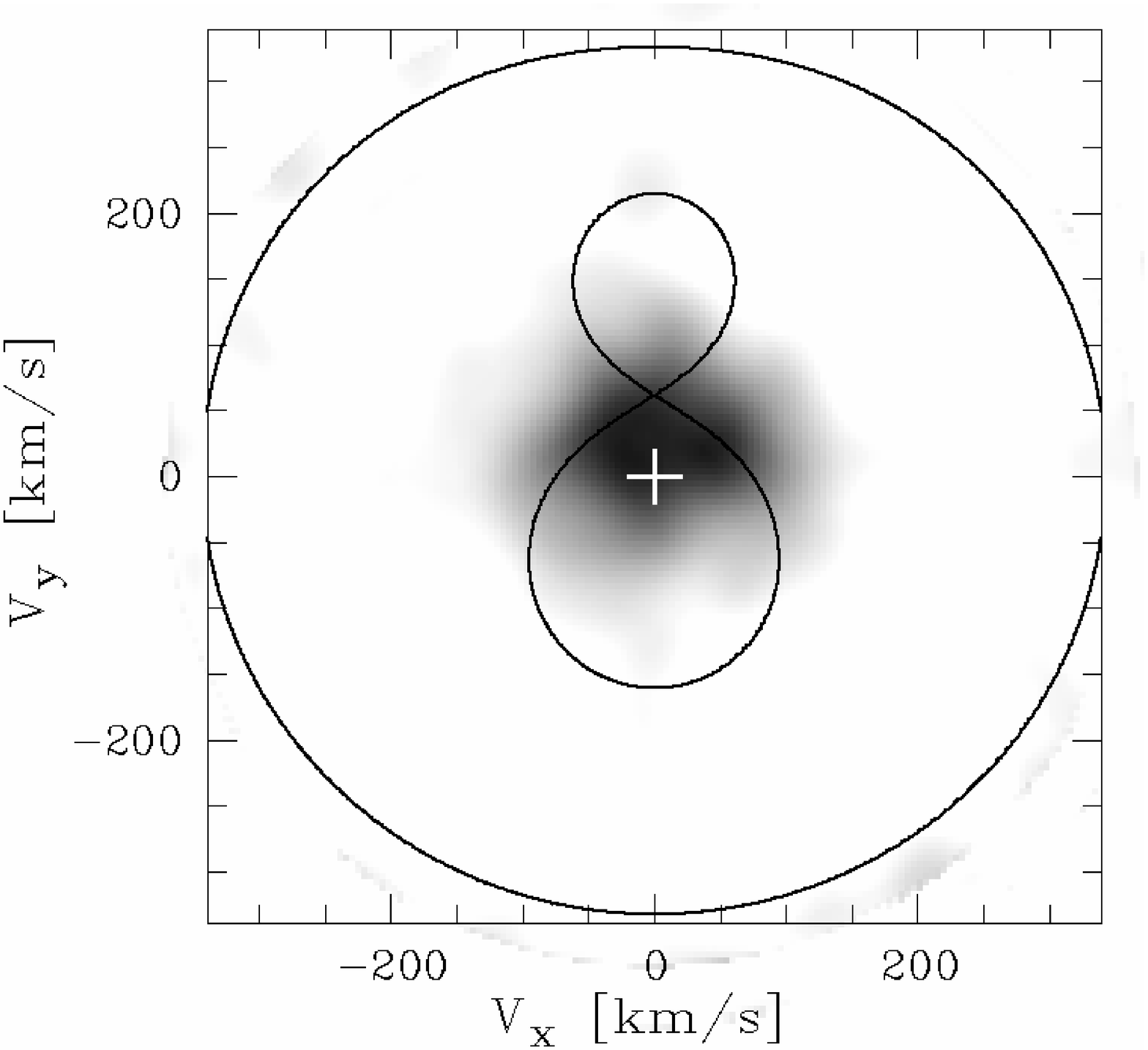}}
\hfill
\resizebox{4.1cm}{!}{\includegraphics{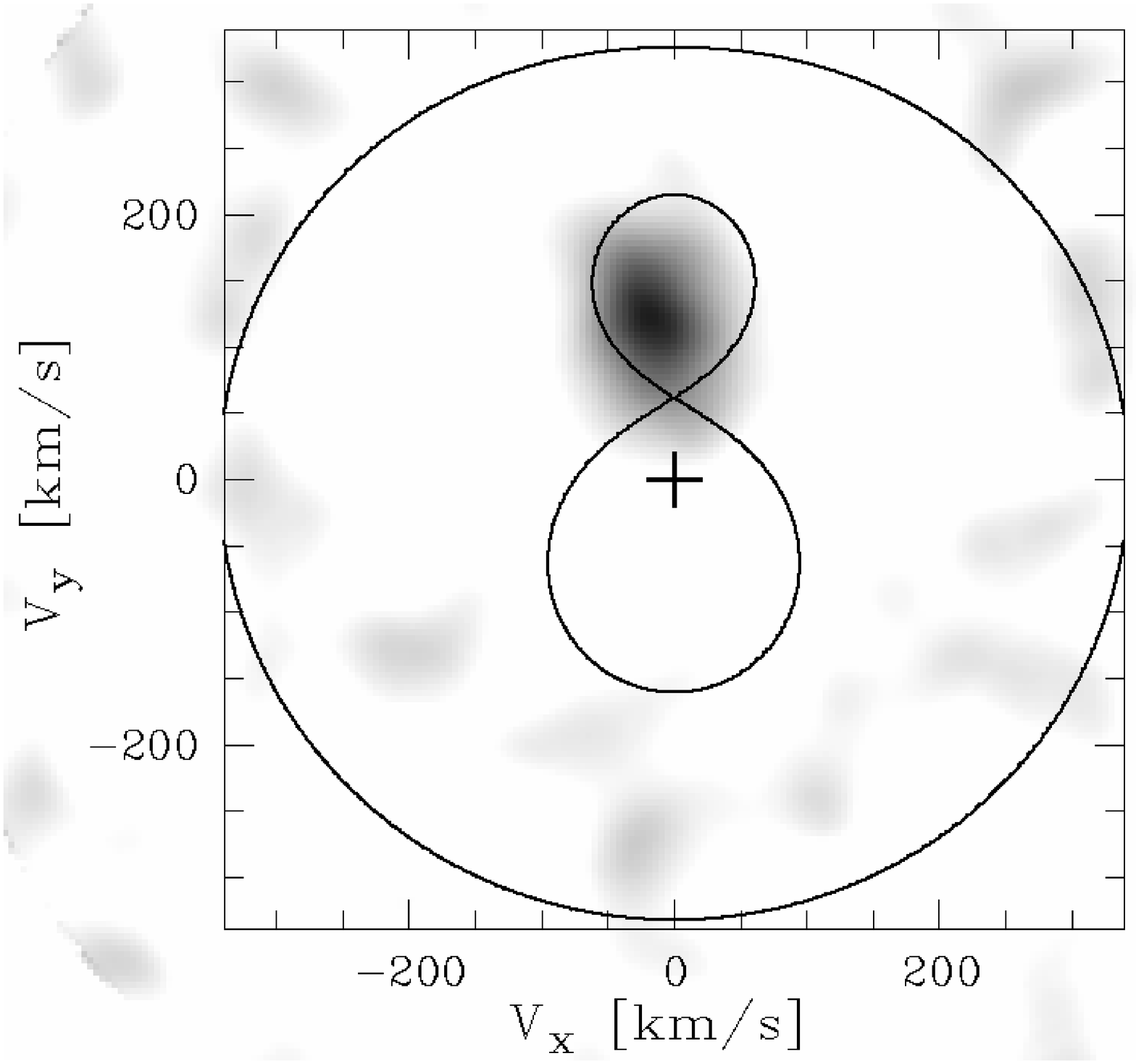}}
\caption{\label{dop_he_na}Doppler maps of He\,I (left) and Na\,D (right), see text for details. 
Overplotted is the Roche geometry for $M_2/M_1 = 0.41$ and $i = 30^\circ$. 
For comparison, the
plot of H$\alpha$ with the same Roche geometry is in Fig.\ \ref{dop_ha}\,b).}
\end{figure}

The different velocity amplitude for the various narrow emission lines can 
be explained by assuming a temperature gradient on the secondary
star as is expected for an irradiated object. He\,I, which needs the 
highest temperature to be excited is only present where the secondary
star is closest to the white dwarf, i.e around L1. 
The temperature to excite hydrogen to emit H$\alpha$ is lower, hence the
irradiated region on the secondary which emits H$\alpha$ is larger and 
due to the bound rotation of the secondary reaches to higher velocities.
The Na lines instead are probably chromospheric and thus trace the motion 
of the secondary's centre of mass, while the TiO absorption bands 
come from
the non-irradiated backside of the secondary. 

For the He and Na emission lines, this behaviour is 
visualised in Fig.\ \ref{dop_he_na}. Here, we computed the Doppler maps for
He\,I\,$\lambda$5878, He\,I\,$\lambda$6678, and He\,I\,$\lambda$7065 individually and averaged 
them to increase the S/N ratio. 
For the Na doublet instead, the two lines were too close for computing
individual Doppler maps. Instead, we
shifted each spectrum by 5.97\,\AA\ -- the wavelength
difference between the two lines -- and averaged it with the non-shifted 
spectrum. This yields a central averaged line and two outer lines of 
half the strength which appear in the Doppler map as an artificial structure
 following an outer ring centred around the Na emission source which 
has to be ignored. It is clear from the Doppler maps in Fig.\ \ref{dop_ha} and
\ref{dop_he_na} that the regions of line emission are different for
each element and qualitatively follow the expected temperature distribution
for an irradiated secondary star if a mass ratio of $q\ge 0.4$ is assumed. 
For smaller mass ratios, the Roche-geometry changes such that the
irradiated material would originate 
rather in the area between L1 and the centre of rotation. 

\section{Conclusions}
We observed BB\,Dor during a low state using medium resolution 
VLT/FORS2 spectroscopy around H$\alpha$.
By comparing the line profiles and the continuum with the results of 
\cite{rodriguez-giletal12} we can conclude that no detectable accretion
happened during this observing run. 

We detect H$\alpha$, He\,I, and Na\,I emission lines as well as TiO absorption 
bands which trace
the motion of different temperature zones of the irradiated secondary star. 
Given that He\,I and H$\alpha$ originate at the higher temperature 
regions of the secondary close to L1 (which is reflected in their low 
velocity amplitude) and TiO bands are likely to originate in the non-irradiated,
cooler, backside zones of the secondary
star, we judge that the velocity measured from the Na\,I lines which are likely
of chromospheric origin reflects best the projected velocity of the
secondary. We thus determine the radial velocity amplitude 
$K_2 = 118 \pm 4\,\rm km\,s^{-1}$ for the secondary star.

In H$\alpha$, we find satellite lines similar to those seen in AM\,Her in quiescence.
Our data show clearly that these are described by two crossing sinusoidal curves 
with the same radial
velocity amplitude of 340\,km\,s$^{-1}$, larger than that of the central line or even than $K_2$. As in the case
of AM\,Her, they are
symmetrically offset by $\pm 0.15$ phases. 

If the H$\alpha$ satellites were emitted by material trapped in the 
L4 and L5 Lagrangian points,
following the standard Roche model the necessary mass ratio would be 0.02. 
Even though this value would actually yield stable equilibria, it does
not agree with the possible masses for the components of a CV with an 
orbital period of nearly four hours. The interpretation of trapping the
material by a pure Roche potential is thus refuted. Additional forces, probably
of magnetic nature, are needed to modify the Roche potential to keep
the observed H$\alpha$-emitting material in a stable configuration.
We point out that the white dwarf in BB\,Dor is not 
strongly magnetic. 
With less than 5\,MG, its 
strength is not comparable to AM\,Her.
Still, the observed satellite lines are strikingly similar. 
We thus conclude that  
they are not likely caused by the magnetic field of the white dwarf as was suggested
for the case of AM\,Her \citep{kafkaetal08-1}. We rather
favour the influence of the magnetic field of the secondary as an explanation
for the two additional emission sources. A modified Roche geometry
taking into account such an additional force might result in stable equilibrium
points around the positions where the increased emission is observed.

The other way of explaining the presence of the satellite lines would be 
for H$\alpha$--emitting material moving in prominences 
originating in a magnetically active secondary star. To explain the observed 
variation of the satellite lines, only two orientations are 
possible for these prominences. In this picture the 
similarities between the satellite lines in different CVs are either 
coincidental or more likely are caused by a preferential orientation 
of prominences in a fast orbiting binary star.

\section*{Acknowledgements}
Partially funded by the Spanish MICINN under the Consolider-Ingenio 2010 Program grants CSD2006-00070: First Science with the GTC and CSD2009-00038: ASTROMOL.
PRG thanks ESO for a stay within the Visitor Scientist program.
The use of {\tt MOLLY} developed 
by Tom Marsh is gratefully acknowledged. 
We thank the team on Paranal for taking the
spectra in ToO mode, i.e. Emanuela Pompei and Thomas Rivinius.
\bibliographystyle{mn2e}

\begin{thebibliography}{}

\bibitem[\protect\citeauthoryear{{Appenzeller}, {Fricke}, {F{\"u}rtig}, {G{\"
  a}ssler}, {H{\"a}fner}, {Harke}, {Hess}, {Hummel}, {J{\"u}rgens},
  {Kudritzki}, {Mantel}, {Meisl}, {Muschielok}, {Nicklas}, {Rupprecht} \& {et
  al.}}{{Appenzeller} et~al.}{1998}]{appenzelleretal98-2}
{Appenzeller} I.,  {Fricke} K.,  {F{\"u}rtig} W.,  {G{\" a}ssler} W.,
  {H{\"a}fner} R.,  {Harke} R.,  {Hess} H.-J.,  {Hummel} W.,  {J{\"u}rgens} P.,
   {Kudritzki} R.-P.,  {Mantel} K.-H.,  {Meisl} W.,  {Muschielok} B.,
  {Nicklas} H.,  {Rupprecht} G.,    {et al.} 1998, Messenger, 94, 1

\bibitem[\protect\citeauthoryear{{Araujo-Betancor}, {Knigge}, {Long}, {Hoard},
  {Szkody}, {Rodgers}, {Krisciunas}, {Dhillon}, {Hynes}, {Patterson} \&
  {Kemp}}{{Araujo-Betancor} et~al.}{2003}]{araujo-betancoretal03-1}
{Araujo-Betancor} S.,  {Knigge} C.,  {Long} K.~S.,  {Hoard} D.~W.,  {Szkody}
  P.,  {Rodgers} B.,  {Krisciunas} K.,  {Dhillon} V.~S.,  {Hynes} R.~I.,
  {Patterson} J.,    {Kemp} J.,  2003, ApJ, 583, 437

\bibitem[\protect\citeauthoryear{{Chen}, {O'Donoghue}, {Stobie}, {Kilkenny} \&
  {Warner}}{{Chen} et~al.}{2001}]{2001MNRAS.325...89C}
{Chen} A.,  {O'Donoghue} D.,  {Stobie} R.~S.,  {Kilkenny} D.,    {Warner} B.,
  2001, MNRAS, 325, 89

\bibitem[\protect\citeauthoryear{{Dermine}, {Jorissen}, {Siess} \&
  {Frankowski}}{{Dermine} et~al.}{2009}]{2009AA...507..891D}
{Dermine} T.,  {Jorissen} A.,  {Siess} L.,    {Frankowski} A.,  2009, A\&A,
  507, 891

\bibitem[\protect\citeauthoryear{{Dunstone}, {Collier Cameron}, {Barnes} \& {Jardine}}{{Dunstone} et~al.}{2006}]{2006MNRAS.373.1308D}
{Dunstone} N.~J., {Collier Cameron} A., {Barnes} J.~R., {Jardine} M., 2006,
MNRAS, 373, 1308

\bibitem[\protect\citeauthoryear{{G\"ansicke}, {Sion}, {Beuermann}, {Fabian},
  {Cheng} \& {Krautter}}{{G\"ansicke} et~al.}{1999}]{gaensickeetal99-1}
{G\"ansicke} B.~T.,  {Sion} E.~M.,  {Beuermann} K.,  {Fabian} D.,  {Cheng}
  F.~H.,    {Krautter} J.,  1999, A\&A, 347, 178

\bibitem[\protect\citeauthoryear{{Godon}, {Sion}, {Barrett}, {Szkody} \&
  {Schlegel}}{{Godon} et~al.}{2008}]{godonetal08-1}
{Godon} P.,  {Sion} E.~M.,  {Barrett} P.~E.,  {Szkody} P.,    {Schlegel} E.~M.,
   2008, ApJ, 687, 532

\bibitem[\protect\citeauthoryear{{Hall} \& {Ramcey}}{{Hall} \&
  {Ramcey}}{1992}]{hall+ramcey92-1}
{Hall} J.~C.,  {Ramcey} L.~W.,  1992, AJ, 104, 1942

\bibitem[\protect\citeauthoryear{{Hellier}}{{Hellier}}{2001}]{hellier_book}
{Hellier} C.,  2001, {Cataclysmic Variable Stars}.
Springer

\bibitem[\protect\citeauthoryear{{Hoard}, {Linnell}, {Szkody}, {Fried}, {Sion},
  {Hubeny} \& {Wolfe}}{{Hoard} et~al.}{2004}]{2004ApJ...604..346H}
{Hoard} D.~W.,  {Linnell} A.~P.,  {Szkody} P.,  {Fried} R.~E.,  {Sion} E.~M.,
  {Hubeny} I.,    {Wolfe} M.~A.,  2004, ApJ, 604, 346

\bibitem[\protect\citeauthoryear{{Horne}}{{Horne}}{1986}]{horne86-1}
{Horne} K.,  1986, PASP, 98, 609

\bibitem[\protect\citeauthoryear{{Jeffries}}{{Jeffries}}{1996}]{jeffries96-1}
{Jeffries} R.~D.,  1996, in {K.~G.~Strassmeier \& J.~L.~Linsky} ed., Stellar
  Surface Structure Vol.~176 of IAU Symposium, p.~461

\bibitem[\protect\citeauthoryear{{Kafka}, {Honeycutt} \& {Howell}}{{Kafka}
  et~al.}{2006}]{kafkaetal06-1}
{Kafka} S.,  {Honeycutt} R.~K.,    {Howell} S.~B.,  2006, AJ, 131, 2673

\bibitem[\protect\citeauthoryear{{Kafka}, {Honeycutt}, {Howell} \&
  {Harrison}}{{Kafka} et~al.}{2005}]{kafkaetal05-1}
{Kafka} S.,  {Honeycutt} R.~K.,  {Howell} S.~B.,    {Harrison} T.~E.,  2005,
  AJ, 130, 2852

\bibitem[\protect\citeauthoryear{{Kafka}, {Howell}, {Honeycutt} \&
  {Robertson}}{{Kafka} et~al.}{2007}]{2007AJ....133.1645K}
{Kafka} S.,  {Howell} S.~B.,  {Honeycutt} R.~K.,    {Robertson} J.~W.,  2007,
  AJ, 133, 1645

\bibitem[\protect\citeauthoryear{{Kafka}, {Ribeiro}, {Baptista}, {Honeycutt} \&
  {Robertson}}{{Kafka} et~al.}{2008}]{kafkaetal08-1}
{Kafka} S.,  {Ribeiro} T.,  {Baptista} R.,  {Honeycutt} R.~K.,    {Robertson}
  J.~W.,  2008, ApJ, 688, 1302

\bibitem[\protect\citeauthoryear{{Kafka}, {Tappert}, {Ribeiro}, {Honeycutt},
  {Hoard} \& {Saar}}{{Kafka} et~al.}{2010}]{2010ApJ...721.1714K}
{Kafka} S.,  {Tappert} C.,  {Ribeiro} T.,  {Honeycutt} R.~K.,  {Hoard} D.~W.,
   {Saar} S.,  2010, ApJ, 721, 1714

\bibitem[\protect\citeauthoryear{{Knigge}, {Araujo-Betancor}, {G{\"a}nsicke}, {Long}, {Szkody}, {Hoard}, {Hynes}  \& {Dhillon}}{{Knigge}
  et~al.}{2004}]{kniggeetal04-1}
{Knigge} C., {Araujo-Betancor} S., {G{\"a}nsicke} B.~T., {Long} K.~S., {Szkody} P., {Hoard} D.~W., {Hynes} R.~I., {Dhillon}, V.~S. 2004, ApJ, 615, L129

\bibitem[\protect\citeauthoryear{{Knigge}, {Baraffe} \& {Patterson}}{{Knigge}
  et~al.}{2011}]{2011ApJS..194...28K}
{Knigge} C.,  {Baraffe} I.,    {Patterson} J.,  2011, ApJS, 194, 28

\bibitem[\protect\citeauthoryear{{Latham}, {Liebert} \& {Steiner}}{{Latham}
  et~al.}{1981}]{lathametal81-1}
{Latham} D.~W., {Liebert} J., {Steiner} J.~E., 1981, ApJ, 246, 919

\bibitem[\protect\citeauthoryear{{Mason}, {Howell}, {Barman}, {Szkody} \&
  {Wickramasinghe}}{{Mason} et~al.}{2008}]{2008A&A...490..279M}
{Mason} E.,  {Howell} S.~B.,  {Barman} T.,  {Szkody} P.,    {Wickramasinghe}
  D.,  2008, A\&A, 490, 279

\bibitem[\protect\citeauthoryear{{Oksala}, {Wade}, {Townsend}, {Owocki},
  {Kochukhov}, {Neiner}, {Alecian} \& {Grunhut}}{{Oksala}
  et~al.}{2011}]{2011MNRAS.tmp.1620O}
{Oksala} M.~E.,  {Wade} G.~A.,  {Townsend} R.~H.~D.,  {Owocki} S.~P.,
  {Kochukhov} O.,  {Neiner} C.,  {Alecian} E.,    {Grunhut} J.,  2011, MNRAS,
  p.~1620

\bibitem[\protect\citeauthoryear{{Rodr{\'{\i}}guez-Gil}, {G{\"a}nsicke},
  {Hagen}, {Araujo-Betancor}, {Aungwerojwit} \& {et
  al.}}{{Rodr{\'{\i}}guez-Gil} et~al.}{2007}]{rodriguez-giletal07-1}
{Rodr{\'{\i}}guez-Gil} P.,  {G{\"a}nsicke} B.~T.,  {Hagen} H.-J.,
  {Araujo-Betancor} S.,  {Aungwerojwit} A.,    {et al.} 2007, MNRAS, 377, 1747

\bibitem[\protect\citeauthoryear{{Rodr\'{\i}guez-Gil}, {Schmidtobreick} \& {et
  al.}}{{Rodr\'{\i}guez-Gil} et~al.}{2012}]{rodriguez-giletal12}
{Rodr\'{\i}guez-Gil} P.,  {Schmidtobreick} L., {Long} K.~S., {G\"ansicke} B.~T., Torres M.~A.~P., Rubio-D\'\i ez M.~M., Santander-Garc\'\i a M.,  2012, MNRAS, 
  submitted

\bibitem[\protect\citeauthoryear{{Rodr{\'{\i}}guez-Gil}, {Schmidtobreick} \&
  {G{\"a}nsicke}}{{Rodr{\'{\i}}guez-Gil} et~al.}{2007}]{2007MNRAS.374.1359R}
{Rodr{\'{\i}}guez-Gil} P.,  {Schmidtobreick} L.,    {G{\"a}nsicke} B.~T.,
  2007, MNRAS, 374, 1359

\bibitem[\protect\citeauthoryear{{Rodr{\'{\i}}guez-Gil}, {Schmidtobreick},
  {Long}, {Shahbaz}, {G{\"a}nsicke} \& {Torres}}{{Rodr{\'{\i}}guez-Gil}
  et~al.}{2011}]{rodriguez-giletal11}
{Rodr{\'{\i}}guez-Gil} P.,  {Schmidtobreick} L.,  {Long} K.~S.,  {Shahbaz} T.,
  {G{\"a}nsicke} B.~T.,    {Torres} M.~A.~P.,  2011, ArXiv e-prints

\bibitem[\protect\citeauthoryear{{Schwarzenberg-Czerny}}{{Schwarzenberg-Czerny%
}}{1989}]{schwarzenberg-czerny89-1}
{Schwarzenberg-Czerny} A.,  1989, MNRAS, 241, 153

\bibitem[\protect\citeauthoryear{{Shafter},{Szkody},{Liebert},{Penning},{Bond},
\& {Grauer}}{{Shafter} et~al.}{1985}]{shafteretal85-1}
{Shafter} A.~W., {Szkody} P., {Liebert} J.,
        {Penning} W.~R., {Bond} H.~E., {Grauer} A.~D., 1985, ApJ, 290, 707

\bibitem[\protect\citeauthoryear{{Spruit}}{{Spruit}}{1998}]{spru98}
{Spruit} H.~C.,  1998, ArXiv Astrophysics e-prints

\bibitem[\protect\citeauthoryear{{Steeghs}, {Horne}, {Marsh} \&
  {Donati}}{{Steeghs} et~al.}{1996}]{1996MNRAS.281..626S}
{Steeghs} D.,  {Horne} K.,  {Marsh} T.~R.,    {Donati} J.~F.,  1996, MNRAS,
  281, 626

\bibitem[\protect\citeauthoryear{{Tappert}, {Mennickent}, {Arenas}, {Matsumoto}
  \& {Hanuschik}}{{Tappert} et~al.}{2003}]{tappertetal03-1}
{Tappert} C.,  {Mennickent} R.~E.,  {Arenas} J.,  {Matsumoto} K.,
  {Hanuschik} R.~W.,  2003, A\&A, 408, 651

\bibitem[\protect\citeauthoryear{{Thorstensen}, {Ringwald}, {Wade}, {Schmidt}
  \& {Norsworthy}}{{Thorstensen} et~al.}{1991}]{thorstensenetal91-1}
{Thorstensen} J.~R.,  {Ringwald} F.~A.,  {Wade} R.~A.,  {Schmidt} G.~D.,
  {Norsworthy} J.~E.,  1991, AJ, 102, 272

\bibitem[\protect\citeauthoryear{{Townsend} \& {Owocki}}{{Townsend} \&
  {Owocki}}{2005}]{2005MNRAS.357..251T}
{Townsend} R.~H.~D.,  {Owocki} S.~P.,  2005, MNRAS, 357, 251

\bibitem[\protect\citeauthoryear{{Townsley} \& {G{\"a}nsicke}}{{Townsley} \&
  {G{\"a}nsicke}}{2009}]{Town+09}
{Townsley} D.~M.,  {G{\"a}nsicke} B.~T.,  2009, ApJ, 693, 1007

\bibitem[\protect\citeauthoryear{{Young}, {Rottler} \& {Skumanich}}
{{Young} et~al.}{1991}]{youngetal91-1}
{Young} A., {Rottler} L., {Skumanich} A., 1991, ApJ, 378, L25

\bibitem[\protect\citeauthoryear{{Yurchyshyn}, {Yashiro}, {Abramenko}, {Wang} \& {Gopalswamy}}{{Yurchyshyn} et~al.}{2005}]{2005ApJ...619..599Y}
{Yurchyshyn} V., {Yashiro} S., {Abramenko} V., {Wang} H., {Gopalswamy} N.,
2005, ApJ, 619, 599

\bibitem[\protect\citeauthoryear{{Zorotovic}, {Schreiber} \&
  {G{\"a}nsicke}}{{Zorotovic} et~al.}{2011}]{2011A&A...536A..42Z}
{Zorotovic} M.,  {Schreiber} M.~R.,    {G{\"a}nsicke} B.~T.,  2011, A\&A, 536,
  A42

\end{thebibliography}

\bsp

\end{document}